\documentclass[fleqn,twoside]{article}
\usepackage{espcrc2}

\usepackage{graphicx}
\usepackage[figuresright]{rotating}

\newcommand{\Sign}{{\rm Sign}}

\newcommand{\AmS}{{\protect\the\textfont2
  A\kern-.1667em\lower.5ex\hbox{M}\kern-.125emS}}

\title{Superconductivity and Chiral Symmetry Breaking
with Fermion Clusters}

\author{Shailesh Chandrasekharan\address[MCSD]
{Department of Physics, Duke University, Box 90305,
	Durham NC 27708, USA}
        \thanks{This work is supported in part by the Department of
		Energy grant DE-FG02-96DR409845.}}
       
\begin{document}

\begin{abstract}

Cluster variables have recently revolutionized numerical
work in certain models involving fermionic variables. This
novel representation of fermionic partition functions is
continuing to find new applications. After describing results from 
a study of a two dimensional Hubbard type model that confirm a 
superconducting transition in the Kosterlitz-Thouless universality 
class, we show how a cluster type algorithm can be devised
to study the chiral limit of strongly coupled lattice gauge 
theories with staggered fermions.

\vspace{1pc}
\end{abstract}

\maketitle

\section{INTRODUCTION}

During the last few years a new class of fermion algorithms have
emerged. The essential progress is a result of our ability to
rewrite certain fermionic partition functions as a sum over 
configurations of bond variables with positive definite weights 
\cite{Cha99.a,Cha01.a}, i.e.,
\begin{equation}
Z \;=\;\sum_{[b]}\;W[b]\;\overline{\Sign}[b]
\label{bpf}
\end{equation}
where $W[b]>0$ is the weight of the bond configuration $[b]$ and
$\overline{\Sign}[b] \geq 0$ is an entropy factor that takes into 
account degrees of freedom other than the bond variables. Typically,
the Pauli principle is encoded in the topology of clusters formed
by lattice sites connected through the bonds. Clusters also carry
a variety of interesting physical information. For example, sizes 
of certain clusters are related to condensates, the squares of the 
sizes of clusters yield susceptibilities. Further, clusters are
useful in building efficient algorithms close to critical points
where the correlation lengths diverge since they allow non-local 
updates with a reasonable acceptance. This property has helped in 
studying critical phenomena in fermionic models with unmatched 
precision. 

\section{SUPERCONDUCTIVITY}
\label{mca}

The recent success of cluster methods in fermionic systems originates
from the Hamiltonian approach. One starts by writing the 
partition function in terms of configurations of fermion occupation 
numbers whose Boltzmann weight is not guaranteed to be positive 
definite. By introducing additional bond variables which help in
the integration over the fermion occupation numbers, the partition 
function of certain models can be rewritten in the form of eq. (\ref{bpf}). 
Configurations with clusters of a specific topology, called merons, 
then yield $\overline{\Sign}[b] = 0$. 

\begin{figure}[t]
\begin{center}
\includegraphics[width=16pc]{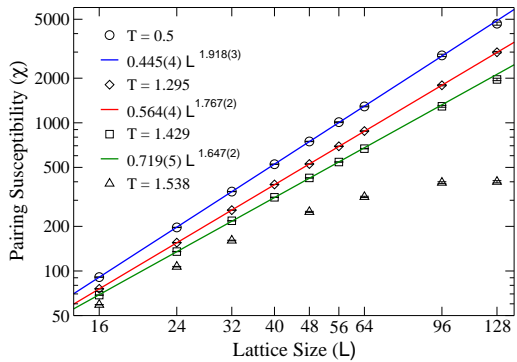}
\caption{ \label{psus} Pairing susceptibility as a function of system size
for various temperatures.}
\end{center}
\vskip-0.3in
\end{figure}

Recently, superconductivity in a two dimensional attractive Hubbard 
type model was studied using the meron cluster approach. The fermion 
pairing susceptibility $\langle \chi\rangle$ is a useful observable.
It is expected to satisfy the finite size scaling formula
\begin{equation}
\langle \chi \rangle \;=\; \left\{ 
\begin{array}{cc}
L^{2-\eta(T)} & T < T_c \cr
\mbox{Const.} & T > T_c
\end{array}\right.
\label{suseq}
\end{equation}
if the superconducting transition belongs to the Kosterlitz-Thouless (KT) 
universality class, with $0 \leq \eta(T) \leq 0.25$,
$\eta(T_c) = 0.25$ and $\eta(0) = 0$. In the specific model studied, 
$\langle \chi\rangle$ turns out to be a sum over the square of the 
size of each cluster in the zero meron sector and the product of the 
size of the merons in the two meron sector. Figure \ref{psus} shows
a plot of the susceptibility as a function of lattice size for various
temperatures. Consistency with KT predictions is clear.

Another observable relevant to the study of superconductivity is the 
spatial fermion winding number susceptibility $\langle W^2\rangle$ . 
Although this is difficult to evaluate with conventional algorithms, 
it is relatively straight forward in the meron cluster approach. 
Each cluster can be assigned a spatial fermion winding number. The 
susceptibility then turns out to be the sum over the square of each 
cluster's spatial winding number in the zero meron sector and the 
product of the spatial winding number of meron clusters in the 
two meron sector. 
In the infinite volume limit below the critical temperature, one can 
combine known results to obtain $2\pi\eta(T) \langle W^2\rangle = 1$.
Results again show consistency with this expectation. Preliminary results 
from this study was presented in \cite{Osb00} and the final analysis 
in \cite{Cha01.b}. 

\section{CHIRAL SYMMETRY BREAKING}

Although the recent success has been applied to Hamiltonian models of 
chiral symmetry breaking \cite{Cha99.b}, cluster methods are applicable to 
more conventional Lagrangian models as well. For example, consider
strongly coupled lattice gauge theory with massless staggered 
fermions in which the $U(1)$ chiral symmetry is expected to be 
broken spontaneously in four dimensions \cite{Wolf}. This result 
was obtained by mapping the massive model into a statistical mechanics 
of monomer-dimer-polymer (MDP) system with positive definite Boltzmann 
weights and extrapolating the results to the chiral limit. Unfortunately, 
as far as we know, it has been difficult to devise algorithms in the 
chiral limit where the systems become constrained. Local metropolis 
updates which can be formulated 
in the massive case become exponentially inefficient in the chiral 
limit. Here we argue that cluster representations of the MDP systems 
yield useful algorithms directly in the chiral limit.

\begin{figure}[htb]
\begin{center}
\includegraphics[width=15pc]{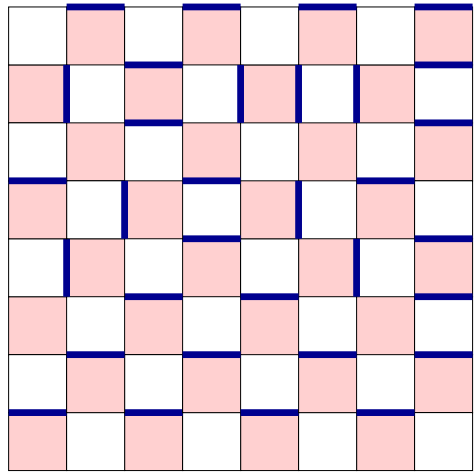}
\caption{ \label{cpdconf} A two dimensional CPD configuration}
\end{center}
\end{figure}

To understand the cluster representation consider for simplicity
the strongly coupled $U(1)$ gauge system. The partition function in
this case is given by the number of closely-packed-dimer (CPD) 
configurations on a lattice. A typical CPD configuration in two 
dimensions is shown in Fig. \ref{cpdconf}.
\vskip-0.3in
Such configurations are also of interest in statistical mechanics 
and play an important role in the solution to the 2-d Ising model 
\cite{Wu}. The chiral symmetry of staggered fermions is manifest
in this representation by the fact that the chiral condensate
vanishes since it is impossible to find a CPD configuration with 
one defect (one site has no dimer lines attached to it).  The chiral 
susceptibility on the other hand is non-zero and proportional to 
the ratio of the total number of CPD configurations with two 
defects (two sites are not connected by dimers) and the partition 
function.

\begin{figure}[htb]
\begin{center}
\includegraphics[width=15pc]{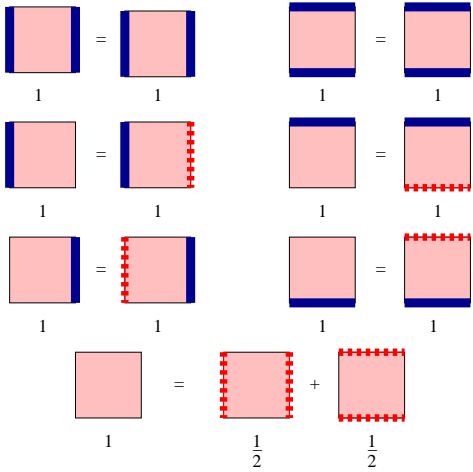}
\caption{ \label{rules} Rules for extending the CPD configurations to
include additional bond variables.}
\end{center}
\vskip-0.3in
\end{figure}

It is possible to extend CPD configurations to configurations of
loops made up of bonds which include the original or ``filled'' dimers 
(represented here by ``solid'' bonds) and ``empty'' dimers (represented by 
``dashed'' bonds) such that the partition function can be 
expressed as a sum over weights of new loop configurations. 
Figure \ref{rules} shows the rules of one such extension in two dimensions.
Each shaded plaquette of the CPD configuration of Fig. \ref{cpdconf} 
carries one of the seven plaquette configurations given on the 
left side of equations in Fig. \ref{rules}. It is easy to check that
all constraints are satisfied if each loop is made up of a repeating 
sequence of filled and empty dimers.
The usefulness of the loop variable is that a dimer system can be
updated by ``flipping'' a loop where filled dimers are emptied and 
vice versa. The acceptance of such a flip is reasonable and 
leads to a useful algorithm. 

\begin{figure}[h]
\begin{center}
\includegraphics[width=15pc]{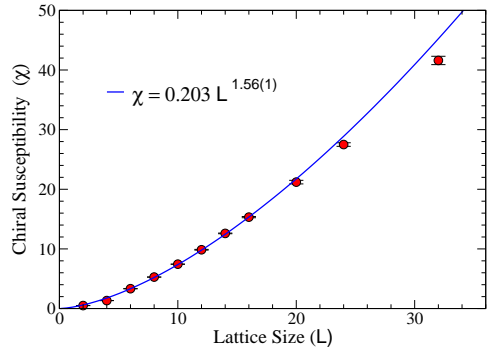}
\caption{ \label{scsus} Chiral susceptibility as a function of lattice size
in two dimensions.}
\end{center}
\vskip-0.3in
\end{figure}

The chiral susceptibility gets contributions when a part of the
loop is flipped and can be measured easily along with the update.
The algorithm was first applied to the two dimensional model.
Although a $U(1)$ chiral symmetry cannot spontaneously break in two 
dimensions, long range correlations can arise as predicted by the 
Kosterlitz-Thouless universality class as discussed in the previous 
section. Figure \ref{scsus} plots the chiral susceptibility with system 
size. Surprisingly, although the data is not inconsistent with the presence
of long range correlations, the susceptibility does not seem to follow 
the predictions of eq. (\ref{suseq}). This puzzle is currently being 
investigated along with extensions to higher dimensions.

I would like to thank J. Osborn and U. Wiese for their collaboration.


\begin{thebibliography}{9}

\bibitem{Cha99.a} S. Chandrasekharan and U.-J. Wiese, Phys. Rev. Lett. 83, 
(1999) 3116; S. Chandrasekharan, Nucl. Phys. (Proc. Suppl.) 83-84,
774 (2000).

\bibitem{Cha01.a} See S. Chandrasekharan, hep-lat/0110018 for a recent 
review.

\bibitem{Osb00} J. C. Osborn, Nucl. Phys. B (Proc. Suppl.) 
Nucl. Phys. Proc. Suppl. 94, (2001) 865.

\bibitem{Cha01.b} S. Chandrasekharan and J. C. Osborn, cond-mat/0109424.

\bibitem{Cha99.b} S. Chandrasekharan, J. Cox, K. Holland and U.-J. Wiese, 
Nucl. Phys. B576, 481 (2000); S. Chandrasekharan and J. C. Osborn, 
Phys. Lett. B496, 122 (2000).

\bibitem{Wolf} P. Rossi and U. Wolff, Nucl. Phys. B248, (1984) 105.

\bibitem{Wu} B. M. McCoy and T. T. Wu, {\em Two Dimensional Ising Model''}, 
Harvard University Press, Cambridge, Massachusetts, 1973.

\end{thebibliography}
\end{document}